\documentclass[cits]{PoS}
\pdfoutput=1
\title{First results of ETMC simulations with $N_{f}=2+1+1$ maximally twisted mass fermions}

\ShortTitle{First results of ETMC simulations with $N_{f}=2+1+1$ maximally twisted mass fermions}

\author{\mbox{R. Baron}$^a$, \mbox{B. Blossier}$^{b}$, \mbox{P. Boucaud}$^{b}$, \mbox{A. Deuzeman}$^{c}$, \mbox{V. Drach}$^{d}$, \mbox{F. Farchioni}$^{e}$, \mbox{V. Gimenez}$^{f}$, \mbox{G. Herdoiza}$^g$, \mbox{K. Jansen}$^g$, \mbox{C. Michael}$^{h}$, \mbox{I. Montvay}$^{i}$, \mbox{D. Palao}$^{f}$, \mbox{E. Pallante}$^{c}$, \mbox{O. P\`ene}$^{b}$, \speaker{S. Reker}\thanks{For the ETM Collaboration} $^{c}$\footnote{Email: {s.f.reker@rug.nl}. Preprint numbers: DESY 09-175, HU-EP-09/50, MS-TP-09-22, SFB/CPP-09-98}, \mbox{C. Urbach}$^{j}$, \mbox{M. Wagner}$^{k}$ and \mbox{U. Wenger}$^{l}$\\
\\\llap{$^a$}CEA, Centre de Saclay, IRFU/Service de Physique Nucl\'eaire, F-91191 Gif-sur-Yvette, France\\
\llap{$^b$}Laboratoire de Physique Th\'eorique (B\^at. 210), Universit\'e de Paris XI, Centre d'Orsay, 91405 Orsay-Cedex, France\\
\llap{$^c$}Centre for Theoretical Physics, University of Groningen, Nijenborgh 4, 9747 AG Groningen, the Netherlands\\
\llap{$^d$}Laboratoire de Physique Subatomique et Cosmologie, 53 avenue des Martyrs, 38026 Grenoble, France\\
\llap{$^e$}Universit\"at M\"unster, Institut f\"ur Theoretische Physik, Wilhelm-Klemm-Stra\ss e 9, D-48149 M\"unster, Germany\\
\llap{$^f$}Dep. de F\'isica Te\`orica and IFIC, Universitat de Val\`encia-CSIC, Dr.Moliner 50, E-46100 Burjassot, Spain\\
\llap{$^g$}NIC, DESY, Platanenallee 6, D-15738 Zeuthen, Germany\\ 
\llap{$^h$}Division of Theoretical Physics, University of Liverpool, L69 3BX Liverpool, United Kingdom\\
\llap{$^i$}Deutsches Elektronen-Synchrotron DESY, Notkestr. 85, D-22603 Hamburg, Germany\\
\llap{$^j$}Helmholtz-Institut f{\"u}r Strahlen- und Kernphysik (Theorie) and Bethe Center for Theoretical Physics, Universit{\"a}t Bonn, 53115 Bonn, Germany\\
\llap{$^k$}Humboldt-Universit\"at zu Berlin, Institut f\"ur Physik, Newtonstra\ss e 15, D-12489 Berlin, Germany\\
\llap{$^l$}Albert Einstein Center for Fundamental Physics, Institute for Theoretical Physics, University of Bern, Sidlerstr. 5, CH-3012 Bern, Switzerland\\}

\abstract{We present first results from runs performed with $N_{f}=2+1+1$ flavours of dynamical twisted mass fermions at maximal twist: a degenerate light doublet and a mass split heavy doublet. An overview of the input parameters and tuning status of our ensembles is given, together with a comparison with results obtained with $N_{f}=2$ flavours. The problem of extracting the mass of the $K$- and $D$-mesons is discussed, and the tuning of the strange and charm quark masses examined. Finally we compare two methods of extracting the lattice spacings to check the consistency of our data and we present some first results of $\chi$PT fits in the light meson sector.}
\FullConference{The XXVII International Symposium on Lattice Field Theory - LAT2009\\
		 July 26-31 2009\\
		 Peking University, Beijing, China}
\begin{document}

\section{Introduction}
The twisted mass formulation of Lattice QCD \cite{Frezzotti:2000nk,Frezzotti:2004wz} is being studied extensively with $N_{f}=2$ dynamical flavours by the European Twisted Mass (ETM) collaboration \cite{Boucaud:2007uk, Boucaud:2008xu,Urbach:2007rt,Blossier:2009bx,Alexandrou:2008tn}. In this formulation of QCD, the Wilson term is chirally rotated within an isospin doublet. To include a dynamical strange quark in a unitary setup, we add, in addition to the strange quark a charm quark in a heavier and mass-split doublet as discussed in \cite{Frezzotti:2003xj,Chiarappa:2006ae,Baron:2008xa}. We will briefly describe our action in section \ref{secaction}, recapitulate our procedure for tuning to maximal twist and focus on the tuning of the heavy doublet. We give an overview of the runs we have carried out and section \ref{secresults} gives first results for some light-quark sector observables.
\section{Lattice setup}\label{secaction}
In the gauge sector we use the Iwasaki gauge action \cite{Iwasaki:1985we}. With this gauge action we observe a smooth dependence of (possible) phase sensitive quantities on the hopping parameter $\kappa$ around its critical value $\kappa_{\rm crit}$. The fermionic action for the light doublet is given by:
\begin{equation}
S_{l}=a^4\sum_x  \left\{\bar{\chi_l}(x)\left[D_W[U] + m_{0,l} + i \mu_l \gamma_5\tau_3  \right] \chi_l(x)\right\},
\end{equation}
using the same notation as used in \cite{Baron:2008xa}. In the heavy sector, the action becomes:
\begin{equation}\label{eqheavyact}
S_{h}=a^4\sum_{x}\left\{\bar{\chi}_{h}(x)\left[D_W[U] + m_{0,h}+i\mu_{\sigma}\gamma_{5}\tau_{1}+\mu_{\delta}\tau_{3}\right]\chi_{h}(x)\right\}.
\end{equation}
At maximal twist, physical observables are automatically $\cal{O}$$(a)$ improved without the need to determine any action or operator specific improvement coefficients. The gauge configurations are generated with a Polynomial Hybrid Monte Carlo (PHMC) updating algorithm \cite{Frezzotti:1997ym,Chiarappa:2005mx,Urbach:2005ji}.
\subsection{Tuning action parameters}\label{subsectunproc}
Tuning to maximal twist requires to set $m_{0,l}$ and $m_{0,h}$ equal to some proper estimate of the critical mass $m_{\rm crit} = m_{\rm crit}(\beta)$ \cite{Frezzotti:2003xj}. Here we set $m_{0,l}=m_{0,h} \equiv 1/(2\kappa)-4$. As has been shown in \cite {Chiarappa:2006ae}, this is consistent with $\cal{O}$$(a)$ improvement defined by the maximal twist condition $am_{{\rm PCAC},l}=0$ (see also ref. \cite{Baron:2008xa}). The numerical precision at which the condition $m_{{\rm PCAC},l}=0$ is fulfilled in order to avoid residual large $\cal{O}$$(a^2)$ effects when the pion mass is decreased is, for the present range of lattice spacings, $|\epsilon/\mu_{l}| \lesssim 0.1$, where $\epsilon$ is the deviation of $m_{{\rm PCAC},l}$ from zero \cite{Boucaud:2008xu, Dimopoulos:2007qy}. As explained in \cite{Baron:2008xa}, tuning to $\kappa_{\rm crit}$ was performed independently for each $\mu_{l}$ value. From table \ref{tabruns} we observe that the estimate of $\kappa_{\rm crit}$ depends weakly on $\mu_{l}$. 
The heavy doublet mass parameters $\mu_\sigma$ and $\mu_\delta$ should be adjusted in order to reproduce the values of the renormalized $s$ and $c$ quark masses. The latter are related to $\mu_\sigma$ and $\mu_\delta$ via \cite{Frezzotti:2003xj}:
\begin{equation}
(m_{s,c})_{\rm R} = \frac{1}{Z_{P}} (\mu_\sigma \mp \frac{Z_{P}}{Z_{S}} \mu_\delta),
\end{equation}
where the $-$ sign corresponds to the strange and the $+$ sign to the charm. In practice we fix the values $\mu_\sigma$ and $\mu_\delta$ by requiring the resulting $K$- and $D$-meson masses to match experimental results.
\begin{table}[h]
\begin{center}
\begin{tabular}{|c|c|c|c|c|c|c|c|c|c|}
\hline
Label & $\beta$ & $\kappa$ & $a\mu_{l}$ & $a\mu_{\sigma}$ & $a\mu_{\delta}$ & $L/a$ & $T/a$ & $m_{\pi}L$ & $|\epsilon/\mu_{l}|$\\
\hline
$C_{1}$ & $1.90$ & $0.1632700$ & $0.0040$ & $0.150$ & $0.190$ & $20$ & $48$ & $3.0$ & 0.14(14) \\
$C_{2}$ & $1.90$ & $0.1632700$ & $0.0040$ & $0.150$ & $0.190$ & $24$ & $48$ & $3.5$ & 0.07(14) \\
$A_{1}$ & $1.90$ & $0.1632650$ & $0.0060$ & $0.150$ & $0.190$ & $24$ & $48$ & $4.1$ & 0.03(3)\\
$A_{2}$ & $1.90$ & $0.1632600$ & $0.0080$ & $0.150$ & $0.190$ & $24$ & $48$ & $4.8$ & 0.02(2)\\
$A_{3}D_{1}$ & $1.90$ & $0.1632550$ & $0.0100$ & $0.150$ & $0.190$ & $24$ & $48$ & $5.3$ & 0.02(2)\\
$A_{4}$ & $1.90$ & $0.1632720$ & $0.0030$ & $0.150$ & $0.190$ & $32$ & $64$ & $4.0$ & 0.08(7)\\
$A_{5}C_{3}$ & $1.90$ & $0.1632700$ & $0.0040$ & $0.150$ & $0.190$ & $32$ & $64$ & $4.5$ & 0.04(5)\\
$A_{6}$ & $1.90$ & $0.1632670$ & $0.0050$ & $0.150$ & $0.190$ & $32$ & $64$ & $5.0$ & 0.05(2)\\
$D_{2}$ & $1.90$ & $0.1632550$ & $0.0100$ & $0.150$ & $0.197$ & $24$ & $48$ & $5.3$ & 0.35(1)\\
$B_{1}$ & $1.95$ & $0.1612400$ & $0.0025$ & $0.135$ & $0.170$ & $32$ & $64$ & $3.4$ & 0.06(6)\\
$B_{2}$ & $1.95$ & $0.1612400$ & $0.0035$ & $0.135$ & $0.170$ & $32$ & $64$ & $4.0$ & 0.02(2)\\
$B_{3}$ & $1.95$ & $0.1612360$ & $0.0055$ & $0.135$ & $0.170$ & $32$ & $64$ & $5.0$ & 0.08(1)\\
$B_{4}$ & $1.95$ & $0.1612320$ & $0.0075$ & $0.135$ & $0.170$ & $32$ & $64$ & $5.8$ & 0.05(1)\\
$B_{5}$ & $1.95$ & $0.1612312$ & $0.0085$ & $0.135$ & $0.170$ & $24$ & $48$ & $4.6$ & 0.01(2)\\
\hline
\end{tabular}
\end{center}
\caption{Input parameters, $m_{\pi}L$ and $|\epsilon/\mu_{l}|$ for all ensembles used in this paper. Every ensemble has $\sim 5000$ thermalized trajectories of length $\tau=1$. We have two main ensemble sets: $A$ and $B$, at $\beta=1.90$ and $\beta=1.95$ respectively. Ensembles labeled $C$ are used to check finite size effects. Ensembles labeled $D$ are used to check/tune the strange and charm quark masses.}
\label{tabruns}
\end{table}
\subsection{Determination of heavy-light meson masses}
Since the twisted mass lattice Dirac operator of the non-degenerate heavy quark doublet (cf. (\ref{eqheavyact})) contains a parity odd and flavour non-diagonal Wilson term, parity as well as flavour are not anymore quantum numbers of the theory. In contrast to parity and flavour conserving lattice formulations, it is not possible to compute correlation functions restricted to a single parity and flavour sector in this setup. While the $K$-meson will remain the lightest state and therefore relatively easy to extract, for a theoretically clean determination of the $D$-meson mass one has to consider the four sectors labeled by parity $\mathcal{P} = \pm$ and $\textrm{flavour} = s/c$ at the same time. And since besides the $K$-meson there are a number of $K + n \times \pi$ states and possibly also ``positive parity $K$ states'' below the $D$-meson, this renders the $D$ a highly excited state. At currently available statistics it seems extremely difficult to extract such a high lying state.

As such, we resort to a different strategy in order to extract this mass. We attempt to determine the mass of the $D$-meson without computing the full low-lying spectrum, e.g.\ we do not determine all low lying states below the $D$. To this end we apply smearing techniques (cf. \cite{Jansen:2008si}, where the same setup was used) to construct highly optimized trial states with large overlap to the $K$- and $D$-meson, and make certain assumptions about these trial states, which will be motivated and detailed in an upcoming publication. To extract the $D$-meson mass, we applied three different methods: (1) solving a generalized eigenvalue problem, (2) performing a multi-exponential fit and (3) rotating the twisted basis correlators back to the physical basis (in order to do this we need to compute the light and heavy twist angle and a ratio of the appropriate renormalization constants). The values of the $D$-meson mass extracted from these three different methods are consistent with each other.
\subsection{Status}
\begin{figure}[h]
\begin{minipage}[ht]{6.9cm}
\includegraphics[width=\textwidth]{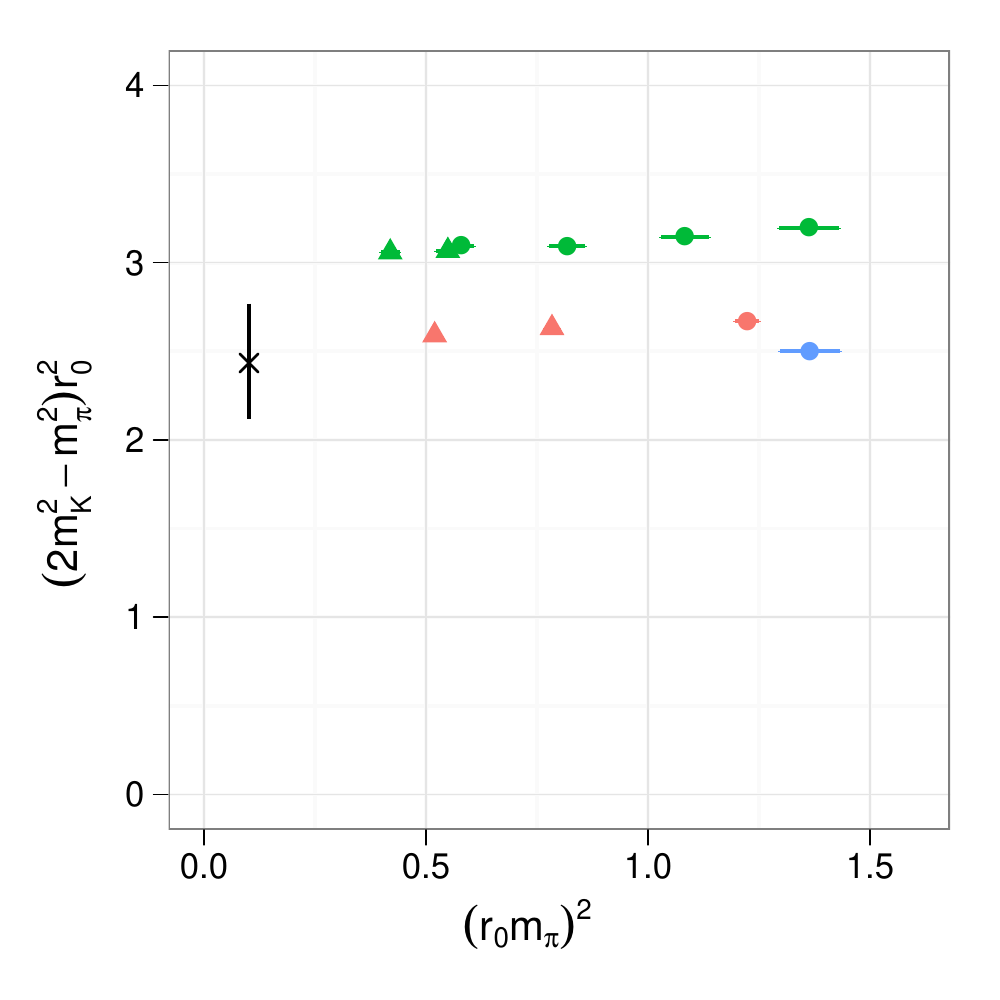} 
\end{minipage}
\hspace{0.5cm}
\begin{minipage}[ht]{8.05cm}
\includegraphics[width=\textwidth]{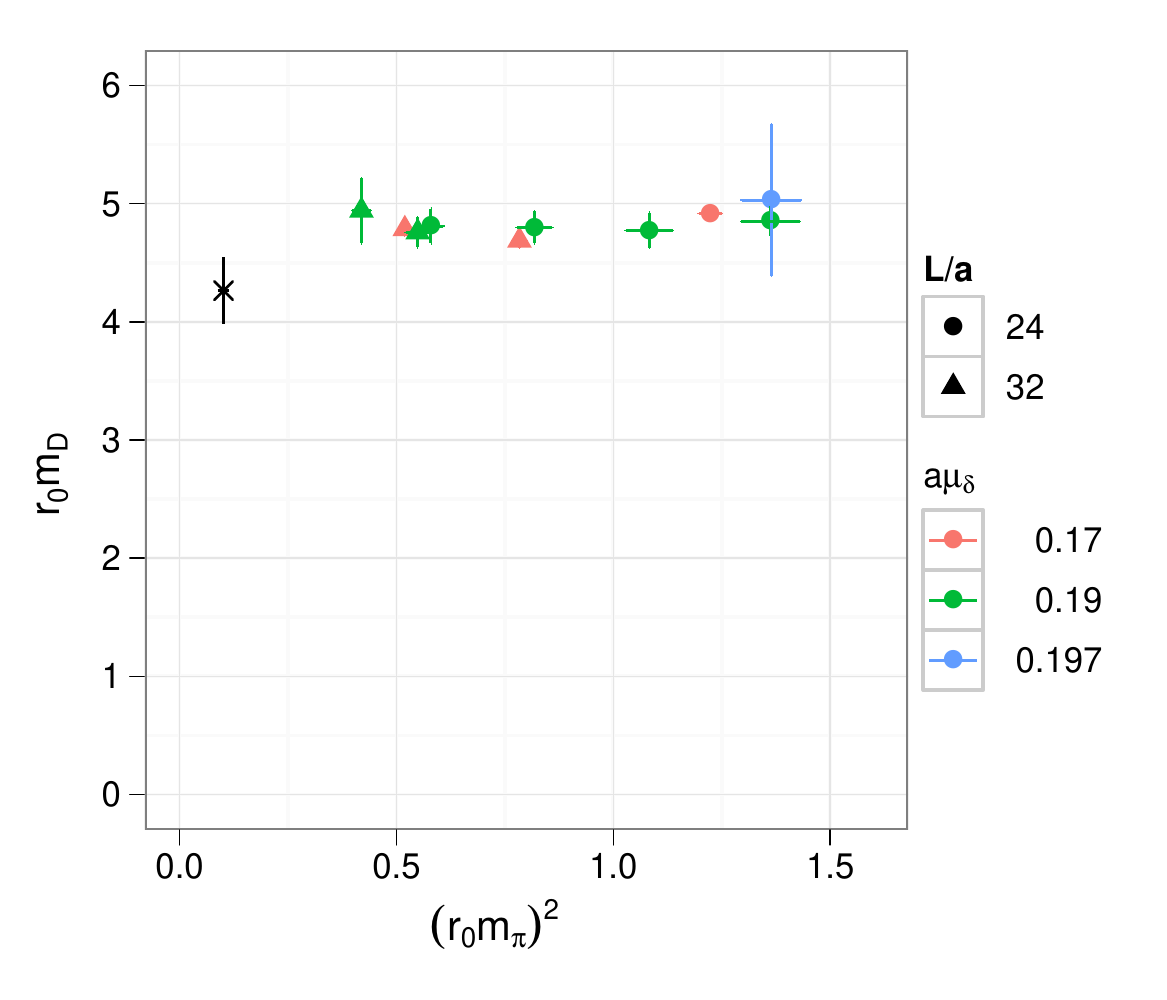} 
\end{minipage}
\caption{$r_{0}^{2}(2m^{2}_{K}-m^{2}_{\pi})$ and $r_{0}m_{D}$ as functions of $(r_{0}m_{\pi})^2$, showing the status of the tuning of the strange and charm quark mass respectively. The experimental value from PDG is added as the black cross ($r_{0}=0.45(3)$ fm was used). Red points label the $\beta=1.95$ runs, green points label the $\beta=1.90$ runs, where the single blue point corresponds to $\beta=1.90$ with a different heavy sector splitting $a\mu_{\delta}$. Circles denote runs with $L/a=24$, triangles indicate a volume with $L/a=32$.}
\label{r0mKfig}\label{r0mDfig}
\end{figure}

The left panel of figure \ref{r0mKfig} shows the tuning of the strange quark mass by showing the difference, scaled with the chirally extrapolated value of $r_{0}/a$ between twice the $K$-meson mass squared and the pion mass squared. Set $A$ at $\beta=1.90$ $a\mu_{\delta}=0.190$ (green points) appears to overshoot the physical point (the black cross on the left), while set $B$ (red points) extrapolates better. To improve the tuning of the strange quark mass for set $A$, we are currently applying a reweighting procedure as described in \cite{Baron:2008xa} in the parameters $a\mu_{\delta}$ and $\kappa$. The blue point with a different heavy sector splitting $a\mu_{\delta}=0.197$ is a run to check this procedure. Though this run is not tuned to maximal twist yet, the $K$-meson mass appears to be much closer to its physical value.
The right panel of figure \ref{r0mDfig} shows the mass of the $D$-meson (obtained in this case by method (3)) as a function of the pion mass squared for various simulation points as well as the experimental value from the Particle Data Group \cite{Amsler:2008zzb} . The plot demonstrates that we have tuned the charm (sea) quark mass in our simulations to a physically realistic value. As a final check, we also use an estimate of $Z_{P}/Z_{S}$ to verify that $m_{c}\sim10m_{s}$.
\section{Results}\label{secresults}
\begin{figure}[h]
\begin{minipage}[ht]{6.9cm}
\includegraphics[width=\textwidth]{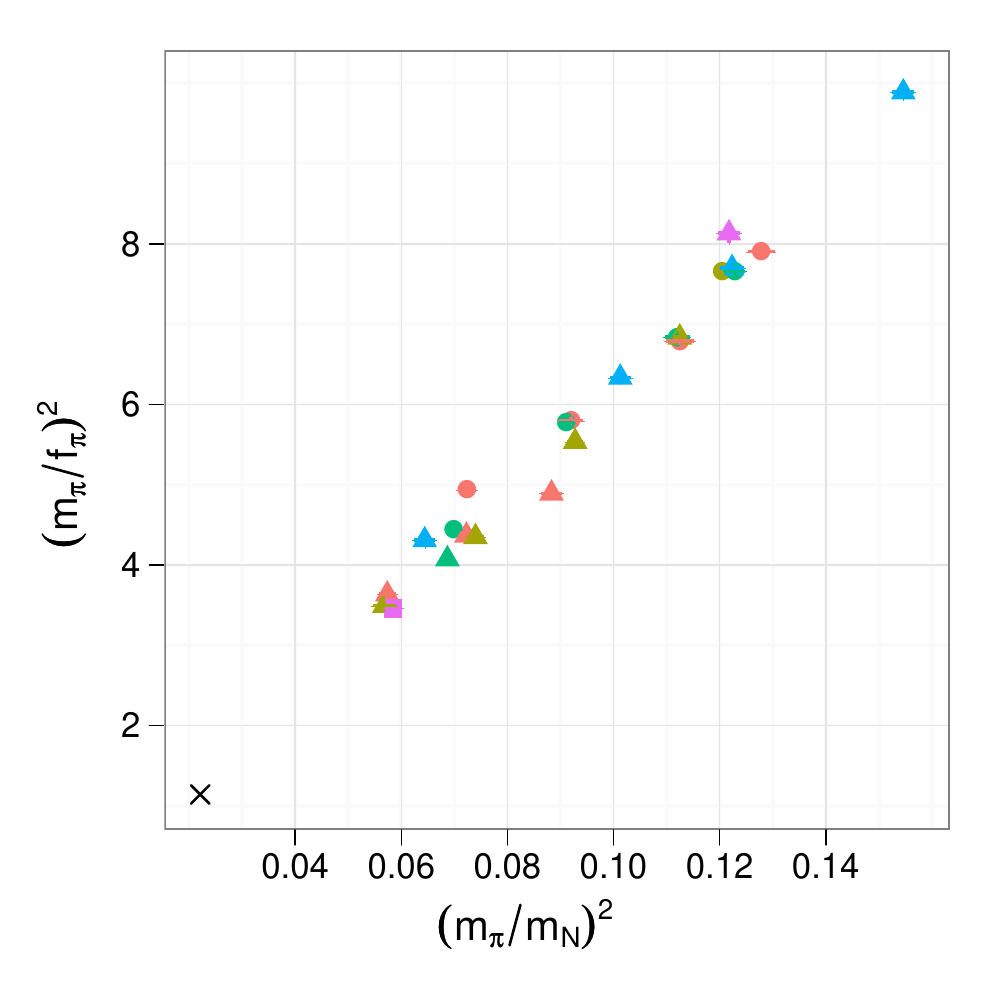} 
\end{minipage}
\hspace{0.5cm}
\begin{minipage}[ht]{8.05cm}
\includegraphics[width=\textwidth]{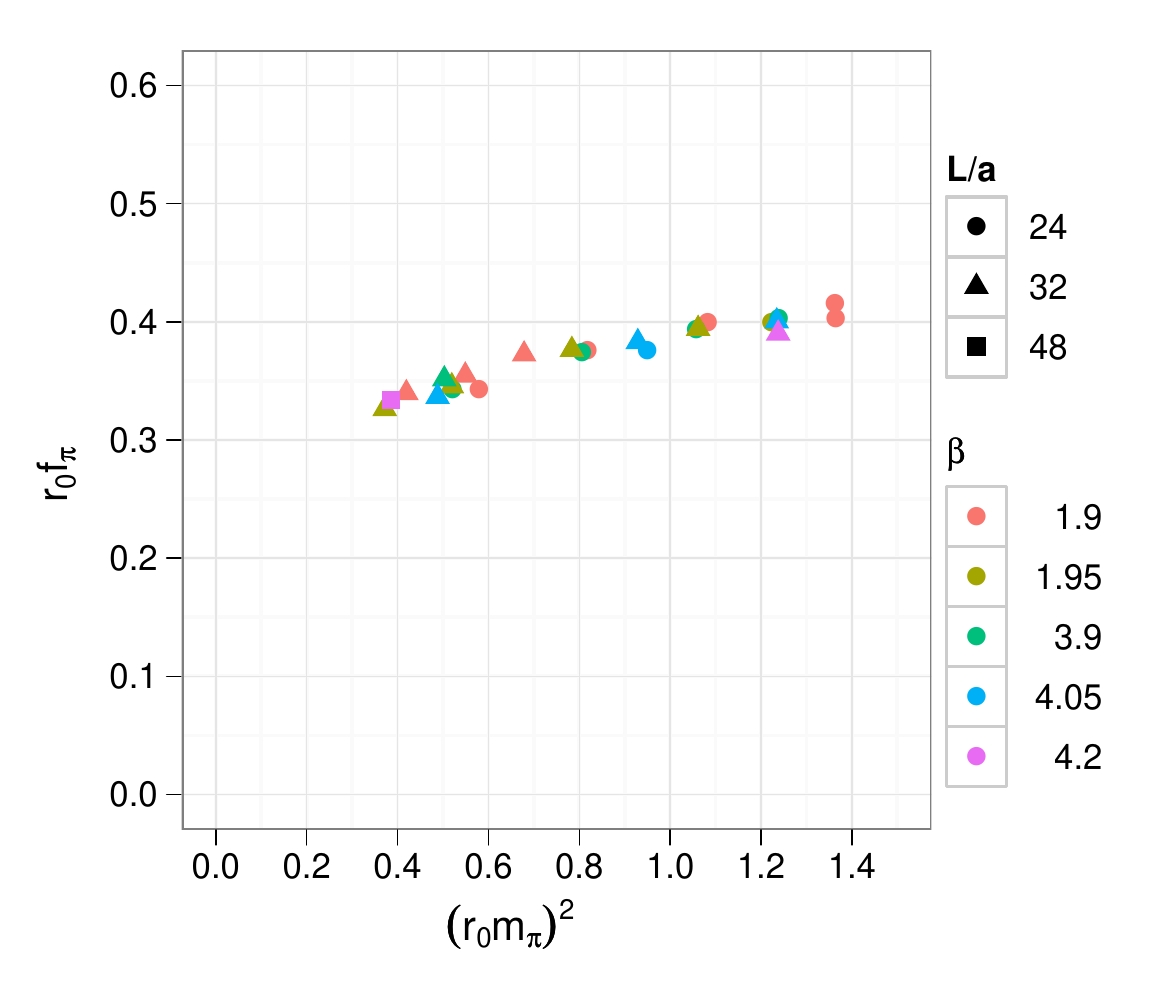} 
\end{minipage}
\caption{$(m_{\pi}/f_{\pi})^2$ vs $(m_{\pi}/m_{N})^2$ (left) and $r_{0}f_{\pi}$ vs $(r_{0}m_{\pi})^2$ (right) for both $N_{f}=2+1+1$ data (with $\beta=1.90,1.95$) and $N_{f}=2$ data (with $\beta=3.9,4.05,4.2$ and using a different gauge action). In the left plot, $m_{N}$ is the nucleon mass, and the physical point is included as the black cross. In both plots finite size corrections are not applied, and in the right plot the chirally extrapolated values for $r_{0}/a$ were used.} 
\label{fpsmnfig}\label{r0fpsfig}
\end{figure}

As a first check of our data, we have compared it to the extensively analysed data set that exists for our $N_{f}=2$ data. To compare the two sets, we plot dimensionless physical ratios in figure \ref{fpsmnfig}. The figure shows no evidence of disagreement between all our results, suggesting small discretisation effects and small effects of dynamical $s$- and $c$-quarks for these observables.

\subsection{Light meson chiral perturbation theory fits}
In order to extract the lattice spacing and light quark mass from our data-sets, we perform a next to leading order $SU(2)$ chiral perturbation theory fit of the $m_{\pi}$ and $f_{\pi}$ data. We use continuum formulae and correct for finite size effects either without any new low energy constants \`a la Gasser and Leutwyler \cite{Gasser:1987zq}, or with $\bar{l}_{1}$ and $\bar{l}_{2}$ added in, as described in \cite{Colangelo:2005gd}. The results are listed in table \ref{tabchipt}. We have performed these fits for ensemble sets $A$ and $B$ separately, and also combined them in a single fit. In table \ref{tabchipt}, we include a systematic error, estimated at $2-5\%$, coming from the dispersion of the values of the fitted parameters between NLO and NNLO. Note that since the quark mass enters the $\chi$PT expression, in order to combine the two sets at different lattice spacings, we need to know the renormalization factor of the quark mass $Z_{\mu}=1/Z_{P}$, a computation which is not yet complete. Assuming that $Z_{P}$ is effectively a function of $\beta$ in the range of parameters we are considering, we can fit the ratio of those $Z_{P}$-values and lattice spacings and extract lattice spacings from the combined fit. In every fit we use as inputs the physical $f_{\pi}$ and $m_{\pi}$, and extract $f_{0}$, $\bar{l}_{3}$ and $\bar{l}_{4}$. A complete analysis (analogous to \cite{Herdoiza}) of the systematic effects is in progress.
\begin{table}[h]
\begin{center}
\begin{tabular}{|c|c|c|c|c|c|c|}
\hline
set & pts & $f_{0}$(MeV) & $\bar{l}_{3}$ & $\bar{l}_{4}$ & $a_{\beta=1.90}$(fm) & $a_{\beta=1.95}$(fm) \\
\hline
$A$ \& $B$ & $11$ & $121(4)$ & $3.5(2)$ & $4.7(2)$ & $0.086(6)$ & $0.078(6)$ \\
$A$ & $6$ & $121(4)$ & $3.4(2)$ & $4.8(2)$ & $0.086(7)$ & \\
$B$ & $5$ & $121(4)$ & $3.7(2)$ & $4.7(2)$ & & $0.078(7)$ \\
\hline
\end{tabular}
\end{center}
\caption{Results from the NLO $SU(2)$ $\chi$PT fits for combined, only set $A$ and only set $B$ respectively. Errors are dominated by a systematic error of $2-5\%$ due to performing an NLO fit. The column "pts" refers to the number of ensembles used in that fit.}
\label{tabchipt}
\end{table}

\subsection{Chiral extrapolation of the nucleon mass}
In this section, we present preliminary results for the light quark mass dependence of the nucleon mass. We consider the one-loop result from heavy baryon chiral perturbation theory (HB$\chi$PT)
\begin{equation}
 m_{N} = m_{N}^{0} -4c_{1}m_{\pi}^2-\frac{3g_{A}^2}{16\pi f_{\pi}^2}m_{\pi}^3
\end{equation}
and fix the scale and light quark mass to the point where the ratio $m_{N}/m_{\pi}$ attains its physical value. We fix $f_{\pi}$ and $g_{A}$ to their physical values ($130.7$ MeV and $1.27$ respectively) as has also been performed in \cite{Alexandrou:2008tn}. Using this procedure, we find a lattice spacing of $0.089(2)$ fm and $0.077(3)$ fm for $\beta=1.90$ and $1.95$ respectively. The $\chi^2/\textrm(d.o.f.)$ of these fits is not very good, and fitting a linear extrapolation appears to be consistent with the data. This is not unique to our data, and has been observed by various collaborations. We therefore perform the linear fit here as well, and absorb the difference between the two extrapolations in the systematic error. A more detailed analysis of the chiral extrapolation of the nucleon mass will be presented in an upcoming study. The lattice spacings that we obtain from the chiral extrapolation of the nucleon mass are $0.089(9)$ fm and $0.077(4)$ fm for set $A$ at $\beta=1.90$ and set $B$ at $\beta=1.95$ respectively.

\subsection{$r_{0}/a$ extrapolation}
Since $r_{0}/a$ is very sensitive to $\kappa$ in the vicinity of $\kappa_{\rm crit}$, the fact that we now tune to maximal twist at every value of $\mu_{l}$, might, with respect to what was done for the $N_{f}=2$ case, in part  provide an explanation for the observed change of slope in the mass dependence of $r_{0}/a$ between $N_f=2$ and $N_f=2+1+1$. Note however that these differences tend to diminish when increasing the value in $\beta$ in the $N_f=2+1+1$ case. We extrapolate $r_{0}/a$ using a simple quadratic fit \mbox{$r_{0}/a = c_{1} + c_{2}a^{2}\mu_{l}^{2}$}, where $c_{1}$ is the value of $r_{0}/a$ in the chiral limit. We perform both a polynomial fit \mbox{$r_{0}/a = c_{1} + c_{2}a\mu_{l}+c_{3}a^{2}\mu^{2}_{l}$} and a linear fit \mbox{$r_{0}/a = c_{1} + c_{2}a\mu_{l}$} to help estimate systematic errors. We find that based on the $\chi^2/\textrm{d.o.f.}$ the quadratic fit is for both values of $\beta$ the best fit. The polynomial fit gives nearly identical results for $c_{1}$, while $c_{1}$ from the linear fit is $1$ to $3\sigma$ higher. Using the lattice spacings from the combined light meson chiral perturbation theory fit, we extract two predictions for $r_{0}$, which seem to agree well at $r_{0}=0.45(3)$ fm.
\begin{table}[h]
\begin{center}
\begin{tabular}{|c|c|c|c|c|c|c|c|c|}
\hline
$\beta$ & $c_{1}$(quadratic fit) & $a$(fm) & $r_{0}$(fm)\\
\hline
$1.90$ & $5.24(2)$ & $0.086(6)$ & $0.45(3)$\\
$1.95$ & $5.71(4)$ & $0.078(6)$ & $0.45(3)$\\
\hline
\end{tabular}
\end{center}
\caption{$r_{0}$ determination for both ensembles separately. $c_{1}$(qua) is the value of a quadratic $r_{0}/a$ extrapolation in the chiral limit with the statistical error in brackets. The lattice spacings $a$ are taken from the combined light meson chiral perturbation theory fit. The obtained values for $r_{0}$ from the two ensembles seem to agree well with each other.}
\label{tabr0a}
\end{table}

\section{Conclusions}
We have presented first results from runs performed with $N_{f}=2+1+1$ flavours of dynamical twisted mass fermions. No evidence of disagreement between these results and those with \mbox{$N_{f}=2$} twisted mass fermions is shown through dimensionless ratio plots (of $m_{\pi}$, $f_{\pi}$, $m_{N}$ and $r_{0}/a$), suggesting small discretisation effects and small effects of dynamical $s$- and $c$-quarks for these observables. We have extracted the lattice spacings of our two ensemble sets using two different methods, which agree within errors with each other. We have measured $r_{0}$ on both ensembles and found consistent results. We are in the process of performing a more detailed combined analysis in order to improve our understanding of the systematic errors. 

\acknowledgments
We thank all other members of the ETM Collaboration for valuable discussions. The HPC resources for this project have been made available by the computer centres of Barcelona, Groningen, J\"ulich, Lyon, Munich, Paris and Rome (apeNEXT), which we thank for enabling us to perform this work. This work has also been supported in part by the DFG Sonderforschungsbereich/Transregio SFB/TR9-03, and by GENCI (IDRIS - CINES), Grant 2009-052271.

\end{document}